%% file: main.tex
\documentclass[sigconf,authorversion,9pt, nonacm, natbib=true]{acmart}

\usepackage{listings}
\usepackage{subcaption}
\usepackage{pifont}
\usepackage[most]{tcolorbox}
\usepackage{multirow}

\input{macros}
\begin{document}
\title{Raqlet: Cross-Paradigm Compilation for Recursive Queries}

\newcommand{\tsc}[1]{\textsuperscript{#1}}
\author{Amir Shaikhha$^{\dagger*}$, Youning Xia$^{\dagger*}$, Meisam Tarabkhah$^{\dagger*}$, Jazal Saleem$^{\dagger}$,Anna Herlihy$^{\ddagger}$}
\affiliation{%
  \institution{$^\dagger$University of Edinburgh, $^\ddagger$EPFL, $^*$Authors contributed equally}
}
\email{amir.shaikhha@ed.ac.uk,y.xia-32@sms.ed.ac.uk,m.tarabkhah@ed.ac.uk,  j.saleem@sms.ed.ac.uk,anna.herlihy@epfl.ch}





\begin{abstract}

We introduce Raqlet, a source-to-source compilation framework that addresses the fragmentation of recursive querying engines spanning relational (recursive SQL), graph (Cypher, GQL), and deductive (Datalog) systems. Recent standards such as SQL:2023's SQL/PGQ and the GQL standard provide a common foundation for querying graph data within relational and graph databases; however, real-world support remains inconsistent across systems. Raqlet bridges this gap by translating recursive queries across paradigms through leveraging intermediate representations (IRs) grounded in well-defined semantics; it translates Cypher or SQL/PGQ to PGIR (inspired by Cypher), then into DLIR (inspired by Datalog), and finally to SQIR (inspired by recursive SQL). Raqlet provides a shared semantic basis that can serve as a golden reference implementation for language standards, while supporting static analysis and transformations (e.g., magic-set transformation) for performance tuning. Our vision is to make Raqlet a robust platform that enables rapid cross-paradigm prototyping, portable recursive queries, and formal reasoning about recursion even when targeting diverse query execution engines.

\end{abstract}

\maketitle




\section{Introduction}

Recursive queries are becoming increasingly important in various data-driven domains. Once considered of limited practical use~\cite{readings1998}, they are now seen as key enablers 
for expressing the complex logic required by modern applications.
Cutting-edge AI-driven knowledge management systems emphasize recursion as a core requirement for expressing rule-based reasoning on graphs~\cite{DBLP:journals/pvldb/BellomariniSG18,DBLP:conf/sigmod/ArefGKLMMMMNPRS25}. In graph analytics and databases, recursion is fundamental for queries such as reachability and shortest paths. In the field of program analysis, deductive databases have emerged as a standard tool for building large-scale static analyzers to encode points-to analyses, dataflow frameworks, and other program analyses~\cite{Smaragdakis2010UsingDF}. Recursive queries are also used in declarative networking to specify network protocols and distributed computations~\cite{dedalus2011}.

Despite this broad importance, support for recursive queries is fragmented across various data systems, each with distinct query languages, capabilities, and performance considerations. In relational databases, SQL has included recursive common table expressions (CTEs) since the SQL:1999 standard. However, recursive CTEs are limited in expressive power, and the query optimization support is restricted to non-recursive subsets. Graph databases offer a specific and limited form of recursion through variable-length path patterns, suitable for reachability and path enumeration. Deductive database systems utilize Datalog, but each Datalog implementation is essentially separate, with unique dialects and optimization strategies~\cite{10.1145/3689767}.

In addition, while relational, graph, and deductive systems all offer support for recursive queries, they do so in distinct and incompatible ways. Key distinctions include data models (tables vs. property graphs vs. logical facts), expressiveness (the types of recursion and logic allowed), and execution strategies (recursive SQL fixpoints vs. pointer-based graph traversal vs. bottom-up Datalog evaluation). These incompatibilities force practitioners to commit to a technology stack early in development, before application requirements are fully understood. Users must either accept the limitations of a high-level language, sacrifice the performance benefits of a specialized engine, or bear the significant cost of migrating between query engines and languages later in development.


Furthermore, despite ongoing efforts to standardize query languages~\cite{Rogova2023ARD}, discrepancies remain in their implementations across different systems~\cite{DBLP:conf/cidr/0001L24}. A ``golden reference implementation'' that is formally specified could help ensure consistency and serve as a common point of reference for various systems.

This paper proposes Raqlet, a novel compilation-based framework that aims to unify these disparate ecosystems for recursive queries. Rather than proposing yet another standalone query engine, Raqlet acts as a source-to-source compiler that translates recursive queries across different languages and systems. 

Raqlet employs intermediate representations (IRs), including Property Graph IR (PGIR), SQL IR (SQIR), and Datalog IR (DLIR), to create a common ground for different query languages. 
For instance, a graph pattern query written in Cypher can be transformed into PGIR, capturing the essential graph traversal semantics. This representation can then be converted to DLIR, a logical rule-based format similar to Datalog with formal fixpoint semantics. Raqlet facilitates seamless bridging between relational, graph, and logic systems, enabling users to write recursive queries in one paradigm and execute them using the optimizations of another, all without the need for manual porting. 

This compilation-oriented approach yields several key benefits.

\smartpara{Static Analysis (Section~\ref{sec:analysis})} Raqlet enables static analysis and reasoning of queries prior to execution. This includes the ability to identify whether recursion is linear or non-linear, verify stratification to prevent illegal negation or aggregation cycles, and prove properties like monotonicity, all of which facilitate more aggressive optimizations. Developers can receive early feedback (e.g., warnings about potential non-terminating recursions or suggestions for query rewrites to improve efficiency) and the system can automatically apply transformations. 

\smartpara{Recursive Optimization (Section~\ref{sec:opt})} Raqlet acts as an optimization catalyst, leveraging extensive query optimization techniques from relational, graph, and logic programming domains on the compiled IR. Methods such as Magic Sets rewriting and linearizing recursive rules can be incorporated within Raqlet’s compilation pipeline, independent of the query’s original language. Raqlet treats recursion as a general computation pattern that can be effectively optimized and executed across various systems.

\smartpara{Formal Semantics (Section~\ref{sec:formal})} Finally, Raqlet establishes formal semantics for all input query languages. The DLIR abstraction is grounded in rigorous logical semantics; it inherits the well-defined semantics of Datalog, ensuring that recursive queries maintain clear and consistent meanings. This applies even to cases where input queries originate from other languages, such as Cypher or SQL.

\input{background}

\section{System Overview}
\label{sec:system}

\begin{figure}[t]
\centering
\includegraphics[width=\columnwidth]{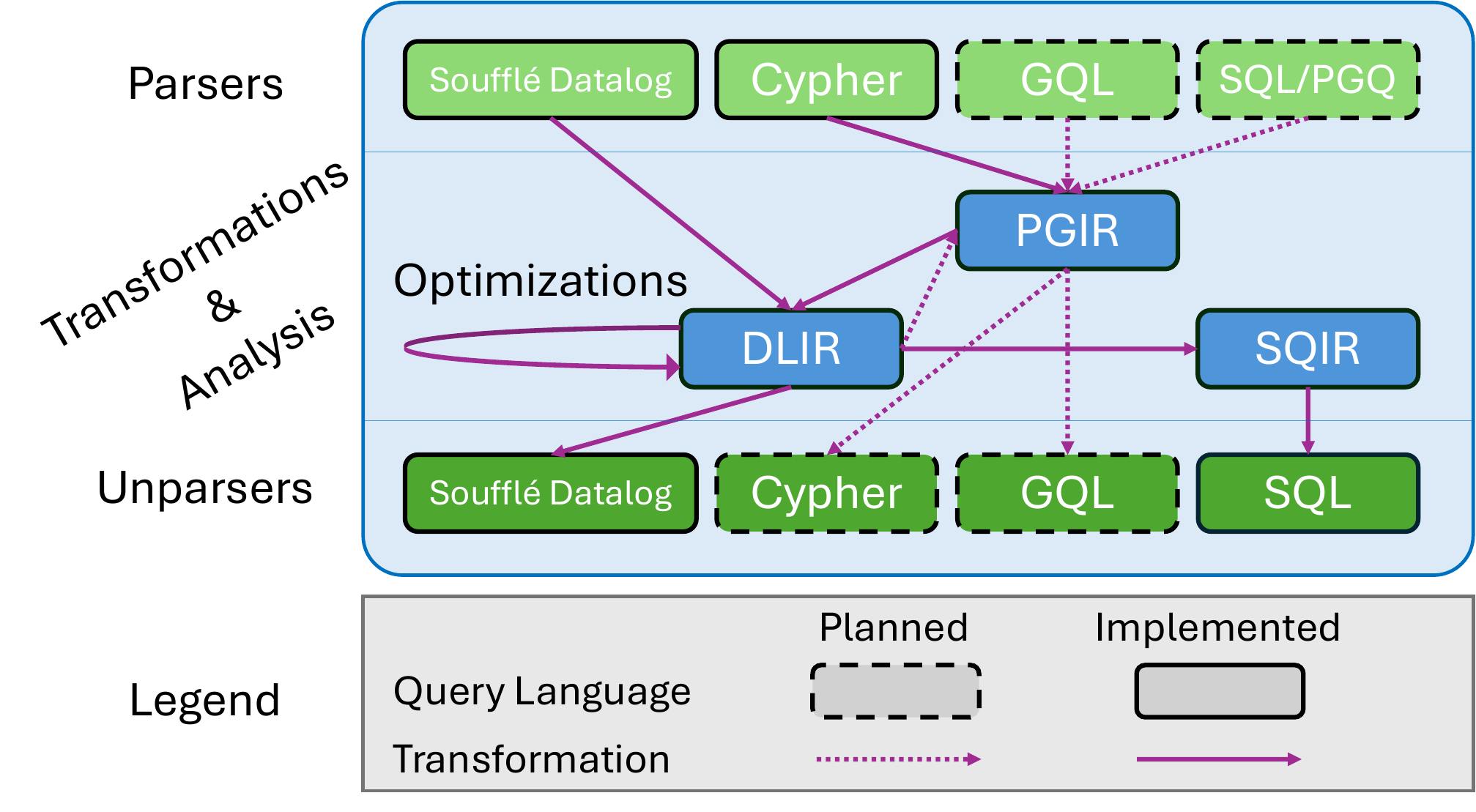}
\vspace{-.5cm}
\caption{Raqlet architecture overview.}
\label{fig:architect}
\vspace{-.3cm}
\end{figure}

Figure~\ref{fig:architect} shows \systemname{}'s architecture. At a high level, \systemname{} involves three main modules: (1) parsers as its frontend, (2) transformations and analysis as its middle end, and (3) unparsers as its backend. 
\systemname{}'s design enables the recursive query to be fully decoupled from its backend-specific representation and lays the foundation of portability across different GDBMSs, RDBMSs, and deductive engines. The key insight is to apply most transformations and analyses at the level of DLIR.
Currently, \systemname{} includes the implementation of a few query languages as input and target languages. We plan to implement the missing frontends and backends in the future. 

We explain the compilation process of the framework through the following running example.

\noindent \textbf{Running Example.} We embed the LDBC Social Network Benchmark (SNB) interactive workload into our environment. Figure~\ref{fig:data-model} presents its schema and Figure~\ref{ex-framework-workflow} shows a simplified version of short query 1~\cite{Erling15}. Given that most deductive databases utilize set semantics and lack certain features such as ordering and limiting the results, to achieve semantic equivalence in translated queries across different backends, we use \code{RETURN DISTINCT} instead and remove \code{ORDER BY} and \code{LIMIT} clauses in input Cypher queries.


\noindent\textbf{Data Model Transformation.} For Cypher queries, \systemname{} takes a PG-Schema $G$ as input and produces DL-Schema, inspired by the Datalog data model, which will be used for query translation. This is done by generating an EDB for every node and edge type in $G$. For instance, the node types \code{personType} and \code{cityType} in Figure~\ref{fig:pg-schema} are translated to EDBs  \code{Person} and \code{City}, respectively, in Figure~\ref{fig:dl-schema}; the edge type \code{locationType} is translated to the EDB \code{Person\_IS\_LOCATED\_IN\_City}.

\noindent \textbf{Cypher to PGIR Translation.} The query translation begins by lowering an input Cypher query into PGIR (Property Graph IR), an intermediate language inspired by GPC~\cite{GPC} but extends to support core Cypher features required for LDBC SNB read workloads, including aggregation and shortest path finding. PGIR represents the query as a sequence of clause constructs such as \code{MATCH}, \code{WHERE}, and \code{RETURN}. The input query undergoes normalization and decomposition into PGIR expressions, such as patterns, filters, and aliasing, which are then mapped to their corresponding clause constructs. 

In our example, the Cypher query (Figure~\ref{ex-cypher}) is passed to the compiler and lowered into its PGIR version. Figure~\ref{ex-pgir} illustrates the graphical representation of PGIR, where grey boxes denote clause constructs, dashed boxes indicate the contents of each clause and arrows imply the order of clauses. 
During lowering, the graph pattern is transformed into PGIR's edge pattern, which consists of edge label \code{IS\_LOCATED\_IN}, a unique identifier \code{x1} generated by the compiler, the type of edge (\code{directed} in this case), and its source and target nodes. Such nodes are represented in PGIR's node pattern which has a node label and an identifier. The transformed edge pattern is mapped to \code{MATCH} construct. Node condition \code{\{id:42\}} is extracted from the graph pattern and mapped to \code{WHERE} construct. Finally, the return statement in Cypher is lowered to \code{RETURN} construct which includes output items. This step simplifies the query representation and hence eases subsequent translations. 

\input{figures/schema_translation}

\input{figures/query_translation}

\noindent \textbf{PGIR to DLIR Translation.} Next, the query is translated from PGIR to DLIR, our core intermediate representation of recursive queries based on Datalog with negation and aggregation. DLIR represents a query as a sequence of rules, with a head specifying an IDB and a body implying how the view is computed. Figure~\ref{ex-dlir} presents the graphical representation of the translated DLIR. Grey boxes denote the head of each DLIR rule, and dashed boxes indicate the body of the respective rules. The atoms that comprise the body are represented by the elements contained within the dashed boxes. This layer of \systemname{} derives semantic-preserving translations for each PGIR construct and employs schema information in Figure~\ref{fig:dl-schema} to deduce variable positions within atoms and infer their types. 

In our example, each PGIR clause construct (i.e., \code{MATCH}, \code{WHERE}, and \code{RETURN}) is translated into a separate DLIR rule (i.e., \code{Match1}, \code{Where1}, and \code{Return}, respectively). The node and edge patterns in PGIR are mapped to their respective EDBs defined in DL-Schema with identifier variables placed at the appropriate position within atoms (note that node id is at the first position of EDB). The node condition \code{n.id = 42} is translated into the predicate \code{n = 42}, and the renaming operation \code{p.id AS cityId} is translated into the variable binding \code{p = cityId}.

\noindent\textbf{DLIR to Datalog and SQL translation.} This layer is responsible for converting DLIR to either Datalog or SQL. Generating Datalog from DLIR is relatively straightforward (Figure~\ref{ex-datalog}). To produce SQL from DLIR, we transform each non-recursive DLIR rule into a Common Table Expression (CTE) and a recursive rule into a recursive CTE. The final SQL program consists of a sequence of CTEs followed by a \code{SELECT} statement from the last CTE.

In the running example, the rules \code{Match1}, \code{Where1}, and \code{Return} are translated into the CTEs \code{V1}, \code{V2}, and \code{V3}, respectively (Figure~\ref{ex-sql}). The logical conjunctions between relation atoms are translated into inner joins, and \code{SELECT DISTINCT} is used to keep set semantics. The final output is selected from the last CTE (\code{V3}). 

\section{Static Analysis and Reasoning}
\label{sec:analysis}
Because recursive query engines vary widely in their expressiveness and evaluation strategies, \systemname{} uses static analysis at the level of DLIR to enable early reasoning about query semantics. DLIR-level static analysis ensures that each analysis is implemented only once, independent of the source query language, and the logical rule-based structure of DLIR makes these analyses straightforward to express and execute.

The key goals of static analyses are (1) identifying unsupported queries by a backend due to semantic limitations, (2) uncovering optimization opportunities that can be expressed with query rewrites, and (3) providing backend-aware guidance to users to identify queries that may cause runtime problems such as non-termination.


\smartpara{Linearity}
Linearity analysis identifies if a query contains linear recursion, which is represented in DLIR with rules that are defined by a single recursive predicate. Linearity is used to reject queries on backends that support only linear recursion, for example RDBMS. \systemname{} can also apply well-known techniques~\cite{10.1109/32.31368} to rewrite non-linear queries into linear ones where possible, to eliminate self-joins and avoid costly materialization of intermediate results.

\smartpara{Mutual Recursion}
Mutual recursion analysis identifies whether a query contains two or more recursive predicates that depend on each other in a cycle. This is used to reject queries on backends that do not support mutual recursion, such as RDBMS. \systemname{} can also use this analysis to enable safe rewrites using existing rewrite techniques for mutually-recursive queries~\cite{lambrecht2025trampoline}.

\smartpara{Monotonicity}
Monotonicity analysis determines whether a recursive query is monotonic under set-inclusion, required by most recursive query engines to guarantee convergence. Non-monotonic constructs, such as negation or certain forms of aggregation, can prevent termination or be rejected at query compile-time by the target execution engine. \systemname{} can use monotonicity analysis to reject unsupported queries and to enable rewrites that safely move aggregates inside recursion when monotonicity is preserved~\cite{10.1145/3299869.3324959}.

\smartpara{Termination Analysis}
Termination analysis identifies whether a recursive query may lead to nontermination at runtime. This includes detecting properties such as interpreted functions over unbounded domains or the use of bag semantics, both of which can lead to infinite recursion~\cite{herlihy2025languageintegratedrecursivequeries}. \systemname{} can use this analysis to warn the user that their queries may not terminate under certain conditions, for example over cyclic data.


\section{Recursive Query Optimization}
\label{sec:opt}
Our framework provides an optimizer to produce more efficient DLIR programs that benefit from well-established as well as novel query optimization techniques.

\smartpara{Inlining} SQL and Datalog queries can involve intermediate views created through CTE clauses or intermediate rules (known as IDBs), respectively. Inlining the definition of such views can not only directly improve the performance, but also indirectly can open up opportunities for further optimizations, including removing self-joins on primary keys~\cite{pytond}.

In our example, Figure~\ref{fig:inline} is an optimized version of Figure~\ref{ex-datalog} where inlining has been
applied. During inlining, rule atoms present in a body (which satisfy certain conditions, i.e. not involved in aggregation and negation) are replaced by their bodies. For instance, \code{Match1} in the body of \code{Where1} is inlined, and so is \code{Where1} in the body of \code{Return}. After inlining, since \code{Person} appears twice (due to a self-join) in \code{Where1}, the duplication is removed.

\input{figures/optimization}

\smartpara{Dead Rule Elimination} In many cases, in particular after inlining rules, there will be intermediate rules that no longer contribute to the final result. In such cases, one can improve the performance by removing such unused rules.

Figure~\ref{fig:dead-rule} further optimizes the previous example by applying dead rule elimination. Since the output rule \code{Return} is not dependent on rules \code{Match1} and \code{Where1}, these two rules are safe to be eliminated from the Datalog program. This will remove unnecessary intermediate results and hence reduce computation complexity. 

\smartpara{Pushing Operators Past Recursion} In non-recursive queries, pushing operators (e.g., selection, projection, or aggregation) past the join is an essential technique to reduce the cost of joins. Similarly, the Datalog literature considers techniques for pushing operators past recursion. This includes optimizations such as the magic-set transformation~\cite{magic-set-1} or FGH rule~\cite{DBLP:conf/sigmod/WangK0PS22}.

\smartpara{Semantic Join Optimizations} Semantic query optimization leverages the integrity constraints to optimize the queries further~\cite{DBLP:journals/tods/ChakravarthyGM90}. The database literature includes techniques for eliminating joins based on reasoning over integrity constraints~\cite{DBLP:journals/tods/MaierMS79,DBLP:journals/tods/AhoBU79}. A novel optimization in the context of property graphs is to eliminate joins on keys that are disjoint, which can be inferred from the knowledge encoded in PG-Schema.

\smartpara{Extensibility and Portability}
Unlike most industrial query optimizers, which are tailored to specific target database systems, our optimizer provides abstractions on the IR level, allowing various optimization options to be easily added or removed, thereby achieving extensibility and portability in query optimization. 

\smartpara{Code Generation} Apart from generating code based on recursive query languages such as Datalog and SQL, \systemname{}'s allows for generating low-level code. This is achieved by introducing additional IRs that include more procedural information on how to run recursive queries~\cite{DBLP:journals/pacmpl/ShaikhhaHSO22,DBLP:journals/pacmmod/ShaikhhaSSN24}. This way, \systemname{} can benefit from the techniques developed for just-in-time query compilation~\cite{DBLP:journals/pvldb/Neumann11,DBLP:conf/sigmod/ShaikhhaKPBD016,lingodb}.

\smartpara{Preliminary Experimental Results} Finally, we evaluate the performance of the translated queries against the original Cypher query, using one of LDBC queries. We study unoptimized and fully optimized versions of the Datalog and SQL queries.
Our experiments are conducted on a system featuring an AMD Ryzen 9 5950X 16-Core Processor operating at 3.4GHz, with 64GB of DDR3 RAM, running Ubuntu 24.04.1 OS. We use Neo4j 5.27.0 to run the original Cypher query and Soufflé 2.4.1 for the Datalog queries. SQL queries are executed using Tableau Hyper API 0.0.21200 and DuckDB 1.1.3.

Table~\ref{tab:result} shows the execution times for LDBC short query 1 and complex query 2 for SF10. In most cases, translated Datalog and SQL queries have lower execution times compared to the original Cypher query. In all cases, the fully optimized versions have better performance compared to the unoptimized queries. We expect similar results for the rest of the queries, but leave a more detailed evaluation for future work.  

\input{figures/results}

\section{Towards Formal Semantics}
\label{sec:formal}

The SQL standard has undergone several revisions since its inception through SQL:1999 (which introduced recursion) to the most recent SQL:2023 (when SQL/PGQ was introduced to support querying graphs). In parallel, the GQL standard~\cite{Rogova2023ARD} provides a formal declarative language for property graph databases that unifies query languages such as Cypher and GSQL. These efforts aim to ensure a consistent formal semantics regardless of the underlying system.

However, despite these standardization efforts, actual systems often deviate in both syntax and semantics.  Existing SQL dialects are inconsistent in recursive semantics~\cite{herlihy2025languageintegratedrecursivequeries} and handling of NULLs~\cite{DBLP:conf/cidr/0001L24}.
Similarly, graph query languages such as Cypher and GSQL continue to evolve independently, introducing features that diverge from the GQL standard specifications. This is due to the lack of a ``golden reference implementation'' that resolves the inconsistencies between the standard specification and system implementations.

\systemname{} addresses this gap by leveraging its Datalog-inspired IR, DLIR, as a formally specified core. DLIR is grounded in the well-known least-fixed-point semantics of stratified Datalog and its extensions, such as Datalog$^o$~\cite{DBLP:conf/sigmod/WangK0PS22}. When a Cypher or SQL/PGQ is compiled to DLIR, it can benefit from a precise logical semantics that follows fix-point logic. This way, DLIR serves as a golden reference implementation for the SQL:2023 and GQL standard.

To further improve this foundation, we plan to formalize DLIR semantics, \systemname{}'s translation pipeline, and DLIR optimizations using proof assistants such as Rocq~\cite{DBLP:journals/pacmpl/AuerbachHMSS17,DBLP:conf/cpp/BegayCM21} (formerly Coq) or Lean~\cite{DBLP:conf/pldi/ChuWCS17}. Inspired by the prior efforts on formalizing SQL semantics to verify the correctness of query optimizers~\cite{DBLP:journals/pacmpl/AuerbachHMSS17,DBLP:conf/pldi/ChuWCS17}, we aim to prove the semantic preservation of transformations from Cypher to SQL/PGQ to DLIR. This way, we ensure a machine-checked semantic core.

\bibliographystyle{ACM-Reference-Format}

\bibliography{refs}

\end{document}

%% file: macros.tex
\definecolor{codeblue}{rgb}{0.28,0.46,0.69}
\definecolor{codegreen}{rgb}{0,0.6,0}
\definecolor{codegray}{rgb}{0.5,0.5,0.5}
\definecolor{codepurple}{rgb}{0.58,0,0.82}
\definecolor{codebrown}{rgb}{0.59,0.29,0}
\definecolor{codeorange}{rgb}{1,0.55,0}
\definecolor{codepink}{rgb}{0.858, 0.188, 0.478}
\definecolor{codeyellow}{rgb}{1, 0.843, 0}
\definecolor{codeteal}{rgb}{0, 0.5, 0.5}
\definecolor{codeindigo}{rgb}{0.294, 0, 0.509}
\definecolor{codered}{rgb}{0.9, 0.17, 0.31}
\definecolor{codecyan}{rgb}{0, 0.6, 0.6}
\definecolor{codeolive}{rgb}{0.5, 0.5, 0.2}
\definecolor{codemagenta}{rgb}{1, 0, 1}
\definecolor{codecerulean}{rgb}{0.16, 0.32, 0.75}
\definecolor{keywordColor}{RGB}{148,0,211}
\definecolor{codenavy}{RGB}{65,105,225}

\newcommand{\systemname}{Raqlet}

\lstdefinelanguage{Cypher}{
  keywords={MATCH, WHERE, WITH, OPTIONAL, AS, RETURN, DISTINCT, AND, NOT, count, sum, input, coalesce},
  keywordstyle=\color{keywordColor}\bfseries,
  ndkeywords={int, string, INT, STRING,uid},
  ndkeywordstyle=\color{black}\bfseries,
  identifierstyle=\color{black},
  sensitive=false,
  comment=[l]{//},
  morecomment=[s]{/*}{*/},
  commentstyle=\color{codegreen}\ttfamily,
  stringstyle=\color{codepurple}\ttfamily,
  morestring=[b]',
  morestring=[b]",
  basicstyle=\fontsize{6}{7}\ttfamily
}

\lstdefinelanguage{pgschema}{
  keywords={CREATE, GRAPH},
  keywordstyle=\color{keywordColor}\bfseries,
  ndkeywords={int, string, INT, STRING,uid},
  ndkeywordstyle=\color{codenavy}\bfseries,
  identifierstyle=\color{black},
  sensitive=false,
  comment=[l]{//},
  morecomment=[s]{/*}{*/},
  commentstyle=\color{codegreen}\ttfamily,
  stringstyle=\color{codepurple}\ttfamily,
  morestring=[b]',
  morestring=[b]",
  basicstyle=\fontsize{6}{7}\ttfamily
}

\lstdefinelanguage{Souffle}{
  keywords={decl, output },
  keywordstyle=\color{keywordColor}\bfseries,
  ndkeywords={number, symbol},
  ndkeywordstyle=\color{codenavy}\bfseries,
  identifierstyle=\color{black},
  sensitive=false,
  comment=[l]{//},
  morecomment=[s]{/*}{*/},
  commentstyle=\color{codegreen}\ttfamily,
  stringstyle=\color{codepurple}\ttfamily,
  morestring=[b]',
  morestring=[b]",
  literate={ :- }{{\textbf{\textcolor{black}{:-}}}}{1}
  { . }{{\textbf{\textcolor{black}{.}}}}{1},
  basicstyle=\fontsize{6}{7}\ttfamily
}

\lstdefinestyle{mystyle}{
    frame=lrtb,
    basicstyle=\ttfamily\footnotesize,
    breakatwhitespace=false,         
    breaklines=true,                 
    captionpos=b,                    
    keepspaces=true,
    numbers=none,
    numbersep=5pt,                  
    showspaces=false,                
    showstringspaces=false,
    showtabs=false,                  
    tabsize=2,
  basicstyle=\fontsize{6}{7}\ttfamily
}

\lstdefinelanguage{SQL}{
  keywords={WITH, RECURSIVE, SELECT, FROM, UNION, DISTINCT, AS, AND, WHERE},
  keywordstyle=\color{keywordColor}\bfseries,
  ndkeywords={},
  ndkeywordstyle=\color{black}\bfseries,
  identifierstyle=\color{black},
  sensitive=false,
  comment=[l]{//},
  morecomment=[s]{/*}{*/},
  commentstyle=\color{codegreen}\ttfamily,
  stringstyle=\color{codepurple}\ttfamily,
  morestring=[b]',
  morestring=[b]",
  basicstyle=\fontsize{6}{7}\ttfamily
}

\newif\ifBlueCode
\BlueCodetrue

\newrobustcmd*\bluecode[1][]
  {%
    \BlueCodetrue
    \mintinline[#1]{markdown}%
  }

\newif\ifRedCode
\RedCodetrue

\newrobustcmd*\redcode[1][]
  {%
    \RedCodetrue
    \mintinline[#1]{markdown}%
  }

\newcommand{\code}[1]{\texttt{\textcolor{black}{#1}}}
\newcommand{\smartpara}[1]{\noindent\textbf{#1}.}

%% file: background.tex
\section{Recursive Query Landscape}
Recursion is supported across a range of systems, each with distinct data models, query languages, and performance considerations. This diversity reflects a fragmented landscape in which no single system offers a definitive advantage across all dimensions of performance and expressiveness. In this section, we survey three major categories of recursive systems and discuss what classes of queries have empirically been shown to perform best on each system.

\subsection{Graph Databases}
\smartpara{Data models} 
Property Graphs (PG) represent graphs as nodes and directed edges. Resource Description Framework (RDF) models graphs as subject-predicate-object triples.
PG supports both node and edge properties, which is nontrivial to model in RDF without 
data duplication, but unlike tables, neither model supports efficient representation of hypergraphs. GDBMS typically offer flexible, schema-optional design, though recent work such as Property Graph Schema (PG-Schema)~\cite{pgschema} provides a formalism for specifying and enforcing strict schemas in PG systems.

\smartpara{Query languages} 
Neo4J~\cite{96b066e5eacd43b299251ec2a7e06a8b} is a widely used PG database with its own query language, Cypher. Cypher has a concise and user-friendly syntax but lacks full expressivity, supporting only 15 of 30 queries in the popular LDBC SNB benchmark~\cite{10.1145/3709722}. 
GQL is an ISO standard~\cite{iso39075} that provides a unified language to facilitate interoperability across graph systems, yet 
it has been shown to be less expressive than recursive SQL~\cite{7c05a7c71e2d44f5a1f6bb6bf0f887a6}.
SPARQL, the W3C standard language for RDF~\cite{Hogan2020}, is a key technology for the semantic web yet is limited in that generalized basic graph patterns (bgps) cannot be expressed in SPARQL~\cite{angles2017foundationsmodernquerylanguages}, but can in Datalog.

\smartpara{Performance} 
Neo4J has been shown to outperform MySQL on certain complex queries~\cite{10.1145/3568562.3568648}.
SPARQL systems have been shown to outperform Neo4J for shortest-path and point-lookup queries, while PostgreSQL performs better than both SPARQL and Neo4J for point-lookup queries yet significantly worse for shortest path~\cite{10.1145/3078447.3078459}.
Benchmarking between GDBMS~\cite{10.14778/3574245.3574270} ultimately shows that different systems excel at different queries.

\subsection{Relational databases}
\smartpara{Data model} 
Graphs can be modeled on RDBMS by relations representing nodes and joins between relations representing edges. As a result, \emph{n}-ary relationships and hypergraphs are easy to model but heavily connected graphs result in many-to-many relations and complex, join-heavy queries~\cite{10.1145/3078447.3078459}.

\smartpara{Query languages} 
SQL's \texttt{WITH RECURSIVE} is based on a restricted form of Datalog, and cannot support non-linear or mutual recursion~\cite{iso9075-1999}. SQL/PGQ is a SQL extension for PG in RDBMS, where graphs are a view of a tabular schema~\cite{iso9075-16-2023} and uses the same pattern matching as GQL.
Both recursive SQL and SQL/PGQ suffer from a lengthy and notoriously difficult-to-read ISO standard specification and lack of complete formal semantics~\cite{DBLP:conf/cidr/0001L24}.

\smartpara{Performance} 
RDBMS have been shown to outperform GDBMS under workloads that make heavy use of group by, sort, and aggregations~\cite{cheng2019}.
DuckDB using \texttt{WITH RECURSIVE} has also been shown to outperform Datalog engines for linear queries without aggregation~\cite{10.1145/3735106.3736533}.

\subsection{Deductive Systems}
\smartpara{Data models} 
Deductive systems model nodes and edges as finite relations of ``facts'', typically with a rigid schema. Facts are stored in the \emph{extensional} database (EDB) and graphs are modeled similarly to RDBMS.

\smartpara{Query languages} 
Queries are expressed using logic-based rules that define how to derive new facts from existing facts. The rules are stored in the \emph{intensional} database (IDB). Deductive database query languages are more expressive than SQL or graph query languages, for example Datalog can express a broad class of queries including mutual and non-linear recursion, and many extensions to Datalog exist to express queries with negation and aggregation. 
While deductive databases lack the same wide usage or commercial support as GDBMS or RDBMS, Datalog has three well-established formal semantic models: model-theoretic, fixed-point, and proof-theoretic that serve as a rigorous foundation for formally reasoning about recursive query behavior.

\smartpara{Performance}
Souffl\'e, a commercial Datalog engine, has been shown to outperform SQLite, PostgreSQL, and Neo4J for classic recursive queries like transitive closure~\cite{DBLP:conf/datalog/BrassW19}.

\subsection{Challenges}

Deciding between database paradigms requires domain expertise, must be done relatively early in application development, and can be costly to revise.
Each system makes trade-offs between data model, expressiveness, and performance, and clear comparisons between systems are rarely straightforward. For example, when comparing expressibility of query languages, SQL with recursion is Turing-complete and can therefore claim to simulate any computation, albeit in a very convoluted and inefficient way. It has been shown that these workarounds, possible in graph query languages as well, are not a replacement for native language support as they can lead to exponential blowup of intermediate results and untenably slow performance~\cite{7c05a7c71e2d44f5a1f6bb6bf0f887a6}. Further, recent empirical results often contradict each other, and results depend heavily on query shape, data structure, and implementation details.
Understanding which queries perform best on which system is sufficiently complex that researchers have applied machine learning to predict the runtime of a transitive closure query across GDBMS, RDBMS, and deductive systems~\cite{DBLP:conf/datalog/BrassW19}. 
While pair-wise translation schemes have been proposed between recursive query languages, for example, SPARQL-to-SQL or SQL-to-Cypher, no existing techniques capture the disparate features of recursive query languages under a single formal semantics.
The fragmentation highlights the need for tools that can capture query functionality across systems and automatically translate between query languages to enable dynamic exploration of the recursive query landscape.

%% file: figures/schema_translation.tex
\begin{figure*}[t]
\centering
\begin{subfigure}{0.52\textwidth}
\begin{lstlisting}[language=pgschema]
CREATE GRAPH { 
  ( personType:Person {id INT, firstName STRING, locationIP STRING}),
  ( cityType:City {id INT, name STRING}),
  (: personType )-[ locationType: isLocatedIn  {id INT} ]->(: cityType )     
}
\end{lstlisting}
\vspace{-0.2cm}
\caption{PG-Schema.}
\label{fig:pg-schema}
\end{subfigure}
\hfill
\begin{subfigure}{0.42\textwidth}
\begin{lstlisting}[language=Souffle]
.decl Person(id: number, firstName: symbol, 
             locationIP: symbol)
.decl City(id: number, name: symbol)
.decl Person_IS_LOCATED_IN_City(id1: number, id2: number,
                                id: number)

\end{lstlisting}
\vspace{-0.2cm}
\caption{DL-Schema.}
\label{fig:dl-schema}
\end{subfigure}
\vspace{-.3cm}
\caption{Schema transformation by \systemname{}. The schema is simplified for presentation purposes.}
\label{fig:data-model}
\vspace{-.3cm}
\end{figure*}

%% file: figures/query_translation.tex
\begin{figure*}
\centering
\begin{subfigure}{0.23\textwidth}
\begin{lstlisting}[language=Cypher]
MATCH
(n:Person 
  {
    id:42
  }
)
  -[:IS_LOCATED_IN]->
(p:City)
RETURN DISTINCT
n.firstName AS firstName,
p.id AS cityId
\end{lstlisting}
\caption{The input Cypher query.}
\label{ex-cypher}
\end{subfigure}
\hfill
\begin{subfigure}{0.3\textwidth}
\centering
\includegraphics[width=0.5\textwidth]{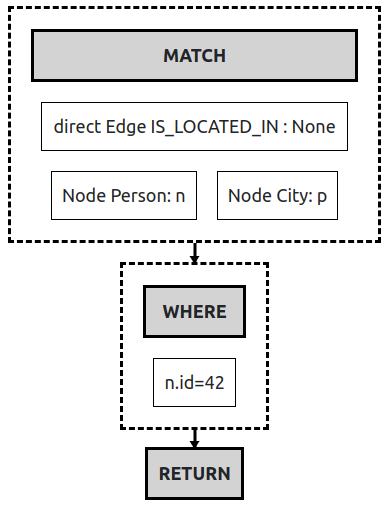}
\caption{PGIR captures the core of Cypher.}
\label{ex-pgir}
\end{subfigure}
\hfill
\begin{subfigure}{0.45\textwidth}
\centering
\includegraphics[width=0.7\textwidth]{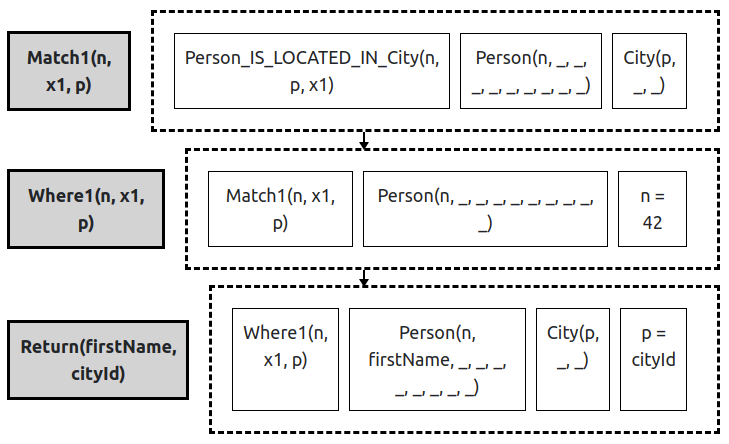}
\caption{DLIR represents the query as a sequence of Datalog rules.}
\label{ex-dlir}
\end{subfigure}


\begin{subfigure}{0.49\textwidth}
\begin{lstlisting}[language=Souffle]
.decl Match1(n: number, x1: number, p: number)
Match1(n, x1, p) :- Person_IS_LOCATED_IN_City(n, p, x1), 
    Person(n, _, _, _, _, _, _, _, _, _), 
    City(p, _, _).
.decl Where1(n: number, x1: number, p: number)
Where1(n, x1, p) :- Match1(n, x1, p), 
    Person(n, _, _, _, _, _, _, _, _, _), 
    n = 42.
.decl Return(firstName: symbol, cityId: number)
Return(firstName, cityId) :- Where1(n, x1, p), 
    Person(n, firstName, _, _, _, _, _, _, _, _), 
    City(p, _, _), 
    p = cityId.
.output Return
\end{lstlisting}
\caption{The generated Datalog program using Soufflè's syntax.}
\label{ex-datalog}
\end{subfigure}
\hfill  
\begin{subfigure}{0.49\textwidth}
\begin{lstlisting}[language=SQL]
WITH V1 AS (
 SELECT DISTINCT R1.id1 AS n, R1.id AS x1, R1.id2 AS p
 FROM Person_IS_LOCATED_IN_City AS R1, Person AS R2, City AS R3
 WHERE (R1.id1 = R2.id) AND (R1.id2 = R3.id)
), V2 AS (
 SELECT DISTINCT V1.n AS n, V1.x1 AS x1, V1.p AS p
 FROM V1, Person AS R1
 WHERE (V1.n = 42) AND (V1.n = R1.id)
), V3 AS (
 SELECT DISTINCT R1.firstName AS firstName, V2.p AS cityId
 FROM V2, Person AS R1, City AS R2
 WHERE (V2.n = R1.id) AND (V2.p = R2.id)
)
SELECT DISTINCT * FROM V3
\end{lstlisting}
\caption{The generated SQL query.}
\label{ex-sql}
\end{subfigure}

\vspace{-.3cm}
\caption{Representations of the example query at different stages of \systemname{}'s translation pipeline.}
\label{ex-framework-workflow}
\end{figure*}

%% file: figures/optimization.tex
\begin{figure}[t]
\centering
\begin{subfigure}{\columnwidth}
\begin{lstlisting}[language=Souffle]
.decl Match1(n: number, x1: number, p: number)
Match1(n, x1, p) :- Person(n, _, _, _, _, _, _, _, _, _), 
  City(p, _, _), Person_IS_LOCATED_IN_City(n, p, x1).
.decl Where1(n: number, x1: number, p: number)
Where1(n, x1, p) :- Person(n, _, _, _, _, _, _, _, _, _), 
  City(p, _, _), Person_IS_LOCATED_IN_City(n, p, x1), n=42.
.decl Return(firstName: symbol, cityId: number)
Return(fn, ci) :- Person(n, fn, _, _, _, _, _, _, _, _), 
  City(p, _, _), Person_IS_LOCATED_IN_City(n, p, x1), n=42, p=ci.
.output Return
\end{lstlisting}
\vspace{-0.2cm}
\caption{Optimization with inlining.}
\label{fig:inline}
\vspace{0.1cm}
\end{subfigure}
\begin{subfigure}{\columnwidth}
\begin{lstlisting}[language=Souffle]
.decl Return(firstName: symbol, cityId: number)
Return(fn, ci) :- Person(n, fn, _, _, _, _, _, _, _, _), City(p, _, _), Person_IS_LOCATED_IN_City(n, p, x1), n=42, p=ci.
.output Return
\end{lstlisting}
\vspace{-0.2cm}
\caption{More optimization with dead rule elimination.}
\label{fig:dead-rule}
\end{subfigure}
\vspace{-0.7cm}
\caption{Examples of optimizations applied to a graph query represented in Datalog.}
\label{fig:optimized-datalog}
\end{figure}

%% file: figures/results.tex
\begin{table}[t]
\caption{Execution time (ms) for each query.}
\label{tab:result}
\vspace{-0.4cm}
\begin{tabular}{c c c c c c}
\hline
            \textbf{Query} & \textbf{Optimized} & \textbf{Neo4j} & \textbf{Soufflé} & \textbf{DuckDB} & \textbf{HyPer} \\ \hline
\multirow{2}{*}{SQ1} & \ding{55} & 72.17 & 0.05    & 24.25   & 0.89  \\ \cline{2-6}
 & \ding{51}   &  -    & 0.02    & 1.78      & 0.78   \\ \hline
\multirow{2}{*}{CQ2} & \ding{55} & 87.85 & 11.70    & 33.18   & 215.85  \\ \cline{2-6}
 & \ding{51}   &  -    & 11.31    & 4.01      & 168.16   \\ \hline
\end{tabular}
\vspace{-0.3cm}
\end{table}